\documentclass{ws-procs9x6-cpt19}
\begin{document}

\newcommand{\refeq}[1]{(\ref{#1})}
\def\etal {{\it et al.}}

\title{A 3+1 Decomposition of the Minimal Standard-Model Extension Gravitational Sector}

\author{Nils A. Nilsson,$^1$ Kellie O'Neal-Ault,$^{2}$, and Quentin G. Bailey$^2$}

\address{$^1$National Centre for Nuclear Research, Pasteura 7, 05-077, Warsaw, Poland}

\address{$^2$Embry-Riddle Aeronautical University, 3700 Willow Creek Road, Prescott, AZ 86301, USA}


\begin{abstract}
    The 3+1 (ADM) formulation of General Relativity is used in, for example, canonical quantum gravity and numerical relativity. Here we present a 3+1 decomposition of the minimal Standard-Model Extension gravity Lagrangian. By choosing the leaves of foliation to lie along a timelike vector field we write the theory in a form which will allow for comparison and matching to other gravity models. 
\end{abstract}

\bodymatter

\section{Introduction}
    Local Lorentz invariance is one of the cornerstones of General Relativity (GR) and modern physics. As such it is an excellent probe of new physics, and Lorentz violation is a large and active area of research.\cite{datatables} The Standard-Model Extension (SME) is an often-used effective field theory framework which includes all Lorentz and CPT violating terms.\cite{colladay1,colladay2,gravity}
        
            The 3+1 (ADM) version of GR is used in for example canonical quantum gravity and numerical relativity.\cite{baumgarte, kiefer}  Here we present a 3+1 decomposition of the minimal SME gravity Lagrangian in the case of explicit Lorentz- symmetry breaking. By choosing the hypersurfaces to be spatial, we write the framework in a form which will allow for comparison and matching to other gravity models.
\begin{figure}[h]
    \begin{center}
        \includegraphics[width=0.7\textwidth]{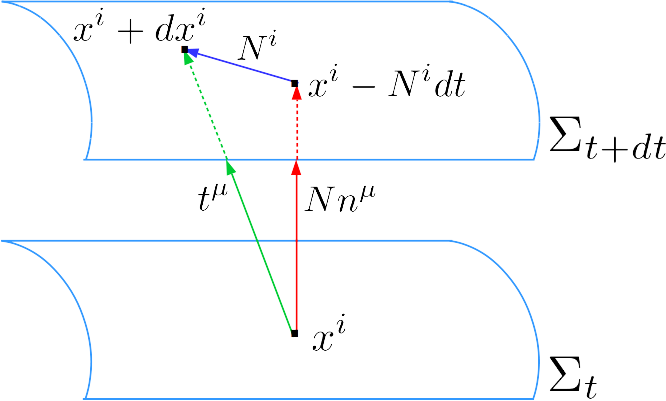}
        \caption{Constant-time hypersurfaces $\Sigma$ along with the ADM variables.}
        \label{fig:adm}
    \end{center}
\end{figure}
\section{The Decomposition}
Using the ADM variables, the metric reads:
\begin{equation}
ds^2 = -N^2 dt^2 + \gamma_{ij}\left(dx^i+N^i dt\right)\left(dx^j+N^j dt\right),
\end{equation}
where $N$ is the lapse function and $N^i$ is the shift vector. These ADM variables relate points on different constant-time hypersurfaces (see Figure~\ref{fig:adm}),
Decomposition of the manifold $\mathcal{M} \rightarrow \Sigma \times \mathbb{R}$ induces the metric $\gamma^{\mu\nu} = g^{\mu\nu} + n^\mu n^\nu$, where $n^\mu=(1/N,-N^i/N)$ is a vector normal to the foliation. The minimal gravitational sector of the SME reads as:\cite{gravity,qb}
\begin{equation}\label{eq:sme}
    \mathcal{L}_{\rm mSME} = \frac{\sqrt{-g}}{2\kappa}\left[-uR + s^{\mu\nu}R^T_{\mu\nu} + t^{\mu\nu\alpha\beta}W_{\mu\nu\alpha\beta}\right],
\end{equation}
where $\kappa=8\pi G$, $R^T_{\mu\nu}$ is the trace-free Ricci tensor, and $W_{\mu\nu\alpha\beta}$ is the Weyl tensor.
In the isotropic limit we can write the above Lagrangian as:
\begin{equation}
\mathcal{L}_{\rm mSME,iso} = \frac{\sqrt{-g}}{2\kappa}\left[\,^{(4)}s_{\mu\nu}\, ^{(4)}R^{\mu\nu}\right],
\end{equation}
where a superscript (4) denotes quantities defined on $\mathcal{M}$. Here, we focus on explicit symmetry breaking so that dynamical terms in the action vanish.\cite{bluhm} The above Lagrangian can be rewritten as:
\begin{multline}
\mathcal{L}_{\rm mSME,iso} = \frac{\sqrt{-g}}{2\kappa}\,^{(4)}s^{\mu\nu}\Big[\gamma^\alpha_{\hspace{4pt}\mu}\gamma^\beta_{\hspace{4pt}\nu}\,^{(4)}R_{\alpha\beta}+n_\mu n_\nu n^\alpha n^\beta \,^{(4)}R_{\alpha\beta} -\\-2\gamma^\alpha_{\hspace{4pt}\mu} n_\nu n^\beta \,^{(4)}R_{\alpha\beta}\Big]
\end{multline}
and by using the Gauss, Gauss-Codazzi, and Ricci equations we can write down the fully decomposed formulation of the gravitational sector (GR + minimal SME):

\begin{multline}
    \mathcal{L}_{\rm EH}+\mathcal{L}_{\rm{mSME, iso}} = \frac{\sqrt{-g}}{2\kappa}\bigg[\left(1+n^\alpha n^\beta \,^{(4)}s_{\alpha\beta}\right)\left(K^{\alpha\beta}K_{\alpha\beta}-K^2\right) +\\+ \left(\frac{1}{3}\,^{(4)}s^i_{\hspace{4pt}i}\right)\mathcal{R} +\nabla_\mu\left(1+\frac{1}{3}\,^{(4)}s^i_{\hspace{4pt}i}-n^\alpha n^\beta \,^{(4)}s_{\alpha\beta}\right)\left(a^\mu+n^\mu K\right)\bigg]
\end{multline}
where $\mathcal{L}_n$ is the Lie derivative along the vector field $n^\mu$, $D_\mu$ is the  covariant derivative associated with the induced metric $\gamma_{\mu\nu}$, $^{(4)}s^i_i$ is the trace of the spatial part of $s^{\mu\nu}$, and $K_{\mu\nu}$ denotes the  extrinsic curvature of the foliation. Moreover, we define the \emph{acceleration vector} $ a_\mu = D_\mu \ln{N}$ and the three-dimensional Ricci scalar $\mathcal{R}$. GR is recovered when $s^{\mu\nu}\rightarrow 0$.
\section{Discussion \& Conclusions}
Using standard tools in numerical relativity theory we have derived a 3+1 decomposition of the minimal SME gravity Lagrangian in the isotropic limit. We make no linearised gravity approximations, and thus this is an \emph{exact} result. This complements other exact studies of the SME.\cite{bonder} Our results can be used in ongoing work on identifying the dynamical degrees of freedom in the explicit symmetry breaking case and matching to proposed models of quantum gravity.
\section*{Acknowledgments}
NAN was partly supported by NCBJ Young Scientist Grant MNiSW 212737/E-78/M/2018.  QGB and KA acknowledge support National Science Foundation grant number 1806871, and support of Embry-Riddle Aeronautical University.

\end{document}